\documentstyle[12pt]{article}
\newcommand{\abs}[1]{\left|#1\right|}

\newcommand{\bra}[1]{\langle #1\vert}
\newcommand{\ket}[1]{\vert #1\rangle}

\newcommand{\be}{\begin{equation}}
\newcommand{\ee}{\end{equation}}

\newcommand{\eg}{e.\,g.\ }
\newcommand{\ie}{i.\,e.\ }

\renewcommand{\Im}{\mbox{\rm Im}}

\renewcommand{\phi}{\varphi}
\renewcommand{\epsilon}{\varepsilon}

\newcommand{\bfnabla}{\mbox{\boldmath $\nabla$}}

\title{Nelsonian Mechanics Revisited\thanks{Submitted to {\em Foundations 
of Physics}.}}
\author{Guido Bacciagaluppi\thanks{Balliol College, Oxford OX1 3BJ, U.K., 
and Sub-Faculty of Philosophy, University of Oxford (e-mail: 
guido.bacciagaluppi@balliol.ox.ac.uk).}}

\date{}

\begin{document}

\maketitle

\begin{abstract}
In de Broglie and Bohm's pilot-wave theory, as is well known, it is 
possible to consider alternative particle dynamics while still preserving 
the $\abs{\psi}^2$ distribution. I present the analogous result for Nelson's 
stochastic theory, thus characterising the most general diffusion processes 
that preserve the quantum equilibrium distribution, and discuss the analogy 
with the construction of the dynamics for Bell's beable theories. I briefly 
comment on the problem of convergence to $\abs{\psi}^2$ and on possible 
experimental constraints on the alternative dynamics.
\end{abstract}

\section{Introduction}\label{introduction}
A well-known feature of de Broglie--Bohm pilot-wave theory (de Broglie, 1928;
Bohm, 1952), first pointed out already by de Broglie ({\em ibid.}), is that
the dynamics preserves the `quantum equilibrium' distribution $\abs{\psi}^2$.
This property does not characterise the dynamics uniquely, however, and
alternative dynamics were discussed by Bohm and Hiley (1993) in the context 
of the Pauli equation.\footnote{The velocity field obtained as the 
non-relativistic limit of the velocity field for the Dirac equation differs 
from the one obtained in the strictly non-relativistic treatment of the Pauli 
equation (Bohm and Hiley, 1993, Sections 10.2 and 10.4). Bohm and Hiley's 
treatment parallels the discussion by Gurtler and Hestenes (1975); see also 
Holland (1993, p. 394). Squires (1996) 
also points out the existence of alternative currents satisfying the 
continuity equation (\ref{six}), with a brief discussion. Finally, in the
stochastic case the possibility of additional velocity terms is 
mentioned by Nelson (1985, p.~55).} Recently, Deotto
and Ghirardi (1998) have given a general discussion of these alternative
dynamics, showing how to construct additional velocity fields satisfying
several important physical constraints.

In this note, I wish to point out that the analogous question can be asked also
in Nelson's stochastic mechanics (Nelson, 1966, 1985) and related theories,
and that Deotto and Ghirardi's results have a direct bearing also to this 
case. I further wish to point out that such a treatment is analogous to that 
of dynamics in BBB (de Broglie--Bohm--Bell) theories (or `beable' theories), 
as discussed by Bell (1984), who constructed such dynamics in the first place
from the requirement that the quantum distribution be preserved in time.
More detailed discussion of such dynamics was given by Vink (1993) and by 
Bacciagaluppi and Dickson (1997).

I briefly review de Broglie--Bohm theory and the Nelson-like stochastic 
theories in Section \ref{configuration}, then summarise Deotto and Ghirardi's 
results for de Broglie--Bohm theory, and present the analogous results for the
stochastic theories in Section \ref{generalisation}. In Section \ref{unique},
following Carlen (1984), I show existence and uniqueness of solutions for the
resulting generalised theories. Section \ref{BBB} spells out the analogy with 
BBB theories. Finally, I conclude with some remarks on
the distribution postulate and possible experimental constraints in Section
\ref{distribution}.

\section{Configuration-Space Theories}\label{configuration}
In de Broglie--Bohm theory, where for simplicity we consider a 
single particle, the state of the system at any time $t$ is given 
by the particle's position ${\bf x}$. The quantum state $\psi$, 
satisfying the Schr\"{o}dinger equation
  \be
    i\hbar\frac{\partial\psi}{\partial t}=
    -\frac{\hbar^2}{2m}\nabla^2\psi + V\psi,
    \label{three}
  \ee
has the role of a {\em pilot wave} or {\em guiding field} for the
particle, and the fundamental equation of motion is de Broglie's
{\em guidance equation}
  \be
    \dot{\bf x}=\frac{\hbar}{m}\bfnabla S,
    \label{four}
  \ee
where $S$ is the phase of the wave function (not divided by $\hbar$).
Straightforward inclusion of a vector potential in (\ref{three}) and 
(\ref{four}) (omitted here) makes the theory gauge invariant.

If a distribution $P({\bf x},t)$ over the position of the particles is given,
we obtain from (\ref{four}) the following expression for a {\em probability 
current}:
  \be
    {\bf j}({\bf x},t)=\frac{\hbar}{m}(\bfnabla S) P({\bf x},t).
    \label{five}
  \ee
And since probability is always conserved, we obtain a {\em continuity 
equation},
  \be
    \frac{\partial P}{\partial t}+\bfnabla\cdot {\bf j}=0,
    \label{six}
  \ee
or
  \be
    \frac{\partial P}{\partial t} +
    \bfnabla\cdot\Big(\frac{\hbar}{m}(\bfnabla S)P\Big)=0.
    \label{seven}
  \ee
It is easy to show that the {\em quantum equilibrium} distribution 
$\abs{\psi}^2$ is a solution of (\ref{seven}), using the Schr\"{o}dinger 
equation (\ref{three}) and writing $\psi=\abs{\psi}\exp (iS)$.

Note that equation (\ref{four}) becomes singular on the set ${\cal N}$ of 
points where $\psi({\bf x})=0$ (the {\em nodal set} of $\psi$). Nonetheless,
as has been shown by Berndl, D\"{u}rr, Goldstein, Peruzzi and Zangh\`{\i}
(1995; see also Berndl, 1996),  the particle has probability zero of entering 
the nodal set from outside. The global existence and uniqueness of solutions 
is then guaranteed for all sufficiently regular initial $\psi$ and all but a 
set of $\abs{\psi}^2$-measure zero initial ${\bf x}$, namely the nodal set 
of $\psi$. 

The condition that the particles be distributed according to $\abs{\psi}^2$
can be taken to imply that the theory is empirically equivalent to
standard quantum mechanics, namely if one argues that all measurement results
in standard quantum mechanics can be described using positions of particles ---
which are all there {\em is} in pilot-wave theory. (The general theory of
measurement in pilot-wave theory was achieved by Bohm (1952, Part II).) 
A question arises thus as to whether there are any
dynamical equations {\em other} than (\ref{four}) that 
preserve $\abs{\psi}^2$.

An example of such a theory is obtained if one modifies
de Broglie--Bohm theory by assuming that the particle
is undergoing not a deterministic evolution but a {\em diffusion process},
so that the dynamics is given by a {\em stochastic guidance equation},
  \be
    d{\bf x} = {\bf b}dt+\sqrt{\alpha}d\mbox{\boldmath $\omega$}   
    \label{twentytwo}
  \ee
(in It\^{o} form), with ${\bf b}=\frac{\hbar}{m}\bfnabla S
+\alpha\frac{\hbar}{2m}\frac{\mbox{\scriptsize \boldmath 
$\nabla$}\abs{\psi}^2}{\abs{\psi}^2}$, and where 
$d\mbox{\boldmath $\omega$}$ is a Wiener process with
  \be
    \overline{d\mbox{\boldmath $\omega$}}=0,\quad 
    \overline{(d\mbox{\boldmath $\omega$})^2}=\frac{\hbar}{m}.
    \label{twentythree}
  \ee
Here, ${\bf b}$ is the {\em drift velocity} of the process and
$\nu:=\alpha\frac{\hbar}{2m}$ is the {\em diffusion coefficent},
treated as a free parameter. Such theories have been discussed by
Bohm and Hiley (1989) under the heading of the {\em stochastic interpretation}
of quantum mechanics, and by Peruzzi and Rimini (1996), who talk about
{\em hidden configurations theories}. Nelson's mechanics 
is recovered if we set $\alpha=1$, and de Broglie--Bohm theory for 
$\alpha=0$.\footnote{This approach differs from that of Nelson and its
generalisation for arbitrary $\alpha$ by Davidson (1979) in that it
postulates a wave function $\psi$ obeying the Schr\"{o}dinger equation.
Nelson's (1966) original paper proposes to {\em derive} the Schr\"{o}dinger
equation from the stochastic particle dynamics. And, as essentially shown
by Davidson (1979) --- but one should use the drift 
velocity ${\bf b}$ instead of his equation (18) --- one can generalise 
Nelson's derivation of the Schr\"{o}dinger equation to the case 
of arbitrary non-zero $\alpha$. However, these derivations of the 
Schr\"{o}dinger equation --- as all derivations based on the supposed 
equivalence with the pair of equations consisting of the continuity 
equation and a Hamilton--Jacobi-type equation --- need to be qualified, 
since they rely on the tacit assumption that the resulting wave function 
be single-valued. As pointed out by Wallstrom (1994), this is a non-trivial 
requirement, which is in fact equivalent to the Bohr--Sommerfeld stability 
condition, \ie to a standard quantisation condition.} 

The conservation equation for such a process can be written as a 
Fokker--Planck equation:
  \be
    \frac{\partial P}{\partial t}+\bfnabla\cdot({\bf b}P-\nu\bfnabla P).
    \label{eight}
  \ee
But now, by substituting $P=\abs{\psi}^2$ one easily sees that (\ref{eight})
reduces to
  \be
    \frac{\partial\abs{\psi}^2}{\partial t}+\bfnabla\cdot
    \Big(\frac{\hbar}{m}(\bfnabla S)\abs{\psi}^2\Big)=0,
    \label{new1}
  \ee
which, as mentioned above, is always satisfied. Thus, $\rho:=\abs{\psi}^2$
is a solution to the Fokker--Planck equation, and, indeed, it can be 
understood as an equilibrium solution, for which the so-called
{\em osmotic current} ${\bf u}P$, where ${\bf u}$ is the {\em osmotic velocity}
  \be
    {\bf u}:=\alpha\frac{\hbar}{2m}\frac{\bfnabla\abs{\psi}^2}{\abs{\psi}^2}
  \ee
and the {\em diffusion current} $\nu\bfnabla P$ balance each other exactly.
The resulting average velocity
  \be
    {\bf v}:= \frac{\hbar}{m}\bfnabla S
  \ee
is called the {\em current velocity}.

Again, when $\psi=0$ the guidance equation becomes singular. On the other 
hand, also in the case of Nelson's mechanics it has been shown 
that under certain regularity conditions on the initial wave function 
and the potential $V$ in the Schr\"{o}dinger equation (\ref{three}), 
the particle will have probability zero of entering the nodal 
set ${\cal N}$, and global existence and uniqueness hold for all initial 
${\bf x}\not\in {\cal N}$ (Carlen, 1984; Nelson, 1985, Sections 11 and 15).
As remarked by Peruzzi and Rimini (1996), these results are valid for 
arbitrary $\alpha$ (cf.\ below, Section \ref{unique}).

Although the dynamics is stochastic, Nelson is careful not to pick out a 
preferred direction in time (in this, he is not followed by Bohm and Hiley).
The time-reversed process is in fact also required to be a diffusion with the
{\em same} osmotic velocity and diffusion coefficient, or equivalently to 
satisfy also the so-called backward Fokker--Planck equation,
  \be
    \frac{\partial P}{\partial t}+\bfnabla\cdot({\bf b}_*P+\nu_*\bfnabla P),
    \label{backward}
  \ee
where
  \be
    {\bf b}_*=\frac{\hbar}{m}\bfnabla S-\nu\frac{\bfnabla\rho}{\rho}
    \quad\mbox{and}\quad
    \nu_*=\nu.
  \ee
Thus,
  \be
    \frac{\partial P}{\partial (-t)}+
    \bfnabla\cdot\Big[(-{\bf v}+{\bf u})P-\nu\bfnabla P\Big],
  \ee
so, indeed, in the time-reversed process only the current velocity changes
sign, the osmotic velocity and diffusion coefficient remaining the same.

\section{Generalisation of the Dynamics}\label{generalisation} 
Deotto and Ghirardi's (1998) question is whether there are any other 
physically reasonable velocity fields ${\bf v}$ apart from (\ref{four})
that yield
  \be
    \frac{\partial\abs{\psi}^2}{\partial t}+\bfnabla\cdot
    ({\bf v}\abs{\psi}^2)=0.
    \label{new3}
  \ee
They point out that, quite obviously, {\em any} velocity field of the form 
${\bf v}=\frac{\hbar}{m}\bfnabla S+{\bf v}_{\scriptsize \rm DG}$
(my notation) with
  \be
    {\bf v}_{\scriptsize \rm DG}=
    \frac{{\bf j}_{\scriptsize \rm DG}}{\abs{\psi}^2},
    \label{new4}
  \ee
where
  \be
    \bfnabla\cdot{\bf j}_{\scriptsize \rm DG}=0,
    \label{new5}
  \ee
also satisfies (\ref{new3}). They then proceed to constrain the extra velocity 
fields by requiring Galilei covariance, the possibility of defining an 
effective wave function and other physically motivated conditions. They also 
note that ${\bf j}_{\scriptsize \rm DG}$ needs to be chosen carefully, so as to
ensure in addition that global uniqueness and existence are still satisfied 
(1998, Section 12). They show that all these constraints can be simultaneously
satisfied and thus show that there are in fact infinitely many deterministic 
pilot-wave-like theories compatible with the condition that the particle 
distributions be given by $\abs{\psi}^2$ at all times.

Deotto and Ghirardi's question can now be modified as follows: what is the
most general {\em diffusion process} (of the form (\ref{twentytwo})) that
preserves the distribution $\abs{\psi}^2$?

Take again the Fokker--Planck equation (\ref{eight}),
  \be
    \frac{\partial P}{\partial t}+\bfnabla\cdot({\bf b}P-\nu\bfnabla P).
    \label{new6}
  \ee
As in Carlen's (1984) proof of global existence and uniqueness, we shall be
concerned with the dynamics on the complement of the nodal set of $\psi$
(and show below that Carlen's results generalise to our case), so that we can
define
  \be
    {\bf w}:={\bf b}-\nu\frac{\bfnabla\abs{\psi}^2}{\abs{\psi}^2}.
    \label{new7}
  \ee
We now solve (\ref{new6})
for ${\bf w}$ with $\abs{\psi}^2$ substituted for $P$:
  \begin{eqnarray}
    {\displaystyle \frac{\partial\abs{\psi}^2}{\partial t} + \bfnabla\cdot
    \Big({\bf w}\abs{\psi}^2 + 
    \nu\frac{\bfnabla\abs{\psi}^2}{\abs{\psi}^2}\abs{\psi}^2 -
    \nu\bfnabla\abs{\psi}^2\Big)    }   &   =   &    \nonumber   \\[1ex] 
    {\displaystyle = \frac{\partial\abs{\psi}^2}{\partial t} + \bfnabla\cdot
    \Big({\bf w}\abs{\psi}^2\Big)   }   &   =   &   0,
    \label{new8}
  \end{eqnarray}
so that
  \be
    {\bf w} = \frac{\hbar}{m}\bfnabla S + {\bf v}_{\scriptsize \rm DG},
  \ee
as above. Thus,
  \be
    {\bf b} = \frac{\hbar}{m}\bfnabla S + {\bf v}_{\scriptsize \rm DG}
              + \nu\frac{\bfnabla\abs{\psi}^2}{\abs{\psi}^2},
    \label{new10}
  \ee
and our desired diffusion process has Fokker--Planck equation
  \be
    \frac{\partial P}{\partial t}+\bfnabla\cdot\Big(
    [\frac{\hbar}{m}\bfnabla S + {\bf v}_{\scriptsize \rm DG}
    + \alpha\frac{\hbar}{2m}\frac{\bfnabla\abs{\psi}^2}{\abs{\psi}^2}]P
    -\alpha\frac{\hbar}{2m}\bfnabla P\Big)=0
    \label{new11}
  \ee
(where we have written again $\nu=\alpha\frac{\hbar}{2m}$), and 
corresponding It\^{o} equation
  \be
    d{\bf x} = [\frac{\hbar}{m}\bfnabla S + {\bf v}_{\scriptsize \rm DG}
    + \alpha\frac{\hbar}{2m}\frac{\bfnabla\abs{\psi}^2}{\abs{\psi}^2}]dt
    +\sqrt{\alpha}d\mbox{\boldmath $\omega$}.
    \label{new12}
  \ee
We see that for ${\bf v}_{\scriptsize \rm DG}=0$ we recover Peruzzi and
Rimini's hidden configuration theories, in particular Nelson's stochastic 
mechanics for $\alpha=1$, while for ${\bf v}_{\scriptsize \rm DG}\neq 0$
and $\alpha=0$ the theory collapses to Deotto and Ghirardi's deterministic 
theories.

Nothing prevents us, however, from interpreting ${\bf v}_{\scriptsize \rm DG}$
as part of the {\em osmotic} velocity rather than the current velocity, or
to split ${\bf w}$ into a ${\bf v}_{\scriptsize \rm DG}$ and a
${\bf u}_{\scriptsize \rm DG}$ that separately (for reasons to become apparent)
satisfy
  \be
    \bfnabla\cdot\Big({\bf v}_{\scriptsize \rm DG}\abs{\psi}^2\Big)=
    \bfnabla\cdot\Big({\bf u}_{\scriptsize \rm DG}\abs{\psi}^2\Big)=0.
    \label{vanish}
  \ee

We also see that if (\ref{new12}) allows for unique global solutions under 
the same conditions as (\ref{twentytwo}) --- in particular, for all ${\bf x}$ 
not in
the nodal set ${\cal N}$ --- then our derivation (in particular (\ref{new7}))
is self-consistent, in the sense that the values of ${\bf w}$ and ${\bf b}$
on the nodal set are irrelevant. We shall now show that under the 
regularity assumptions needed for the standard proof of
global existence and uniqueness in Nelson's mechanics (Carlen, 1984), 
the following further condition is sufficient:
  \be
    \int_s^t\!\int\,
    \abs{{\bf v}_{\scriptsize \rm DG}}^2\abs{\psi}^2d{\bf x}dt' < \infty
    \quad\mbox{and}\quad
    \int_s^t\!\int\,
    \abs{{\bf u}_{\scriptsize \rm DG}}^2\abs{\psi}^2d{\bf x}dt' < \infty
    \label{existence}
  \ee
for any finite time interval $[s,t]$ (cf.\ Carlen, 1984, p. 298; Nelson, 1985,
p. 57). Deotto and Ghirardi's example of a ${\bf v}_{\scriptsize \rm DG}$
for which in the deterministic case global existence and uniqueness hold 
(1998, Section 12) satisfies (\ref{existence}), so that we can use it
to construct an explicit example also for the stochastic case.

\section{Existence and Uniqueness of Solutions}\label{unique}
Carlen (1984) shows that the Fokker--Planck equations (\ref{eight}) and
(\ref{backward}), with $\nu=\frac{\hbar}{2m}$, allow for unique global 
solutions if the following two conditions hold:
  \be
    \int_s^t\!\int\,(\abs{{\bf u}}^2+\abs{{\bf v}}^2)\rho d{\bf x}dt'<\infty,
    \label{car1}
  \ee
and, for any bounded $f$ with bounded continuous first derivatives,
$\int f({\bf x})\rho d{\bf x}$ is differentiable for almost all $t$ and
  \be
    \frac{\partial}{\partial t}\int f({\bf x})\rho d{\bf x}=
    \int(\bfnabla f)\cdot{\bf v}\rho d{\bf x}
    \label{car2}
  \ee
for almost all $t$. Here, $\rho=\abs{\psi}^2$, ${\bf b}={\bf v}+{\bf u}$,
${\bf b}_*={\bf v}-{\bf u}$. (\ref{car2}) yields an interpretation of 
${\bf v}$ as the current velocity, because it is a weakened form of the 
continuity equation
  \be
    \frac{\partial\rho}{\partial t}+\bfnabla\cdot\Big({\bf v}\rho\Big)=0,
  \ee
obtained using $f$ as a test function and integrating by parts. With ${\bf u}=
\frac{\hbar}{2m}\frac{\mbox{\scriptsize \boldmath $\nabla$}\rho}{\rho}$ 
and ${\bf v}=\frac{\hbar}{m}\bfnabla S$ and under the approprate regularity 
assumptions for $\psi$, (\ref{car1}) and (\ref{car2}) are shown to hold.
(And, as remarked by Peruzzi and Rimini (1996), the generalisation to 
arbitrary $\nu$ is straightforward.)

We thus need to show that (\ref{car1}) and (\ref{car2}) are satisfied
also when we take ${\bf u}=
\nu\frac{\mbox{\scriptsize \boldmath $\nabla$}\rho}{\rho}+
{\bf u}_{\scriptsize \rm DG}$ and
${\bf v}=\frac{\hbar}{m}\bfnabla S+{\bf v}_{\scriptsize \rm DG}$.

Indeed, (\ref{car2}) is obviously satisfied, since by partial integration,
  \begin{eqnarray}
    \int\Big(\bfnabla f\Big)\cdot{\bf v}\rho d{\bf x}   & = &
    -\int f\bfnabla\cdot\Big({\bf v}\rho\Big) d{\bf x}= \nonumber \\[1ex] & = &
    -\int f\Big[\bfnabla\cdot\Big(\frac{\hbar}{m}\bfnabla S\rho\Big)+
    \bfnabla\cdot\Big({\bf v}_{\scriptsize \rm DG}\rho\Big)\Big]d{\bf x},
  \end{eqnarray}
and the second term in the integrand vanishes because of (\ref{vanish}).
Further, (\ref{car2}) is satisfied, because
  \begin{eqnarray}
    \int_s^t\!\int\,(\abs{{\bf u}}^2+\abs{{\bf v}}^2)\rho d{\bf x}dt'    & = &
    \int_s^t\!\int\,\Big[\Big(\nu\frac{\bfnabla\rho}{\rho}\Big)^2
                   +\Big(\frac{\hbar}{m}\bfnabla S\Big)^2\Big]\rho d{\bf x}dt'+
    \nonumber  \\[1ex]  & &
    +2\int_s^t\!\int\,\nu\frac{\bfnabla\rho}{\rho}
    \cdot{\bf u}_{\scriptsize \rm DG}\rho d{\bf x}dt'+  \nonumber  \\[1ex] & &
    +2\int_s^t\!\int\,\frac{\hbar}{m}\bfnabla S
    \cdot{\bf v}_{\scriptsize \rm DG}\rho d{\bf x}dt'+
    \nonumber  \\[1ex]  & &
    +\int_s^t\!\int\,\abs{{\bf u}_{\scriptsize \rm DG}}^2\rho d{\bf x}dt'+
    \nonumber  \\[1ex]  & &
    +\int_s^t\!\int\,\abs{{\bf v}_{\scriptsize \rm DG}}^2\rho d{\bf x}dt',
  \end{eqnarray}
where the first integral is finite if Carlen's regularity conditions
are satisfied, the second and third are shown to vanish by partial
integration using (\ref{vanish}) (notice that 
$\frac{\mbox{\scriptsize \boldmath $\nabla$}\rho}{\rho}=\bfnabla\log\rho$), 
and the last two are finite by assumption (\ref{existence}).
This establishes the desired result.

\section{BBB Theories}\label{BBB}
BBB theories are the discrete and stochastic analogue of de Broglie--Bohm
theory. That is, they are theories in which the state of the system
is given by some eigenprojection $P_i$ of some observable $R$,
representing possession of the $i$th eigenvalue. Thus, 
$R$ is a {\em beable} rather than an observable, in Bell's (1987, {\em passim})
terminology. Beable theories have been championed by Sudbery (1986, 1987) and 
have been proposed as a general framework for interpreting quantum mechanics
in Bub (1997), which is the first systematic exposition of beable theories 
in book form.\footnote{Bell's (1984) original proposal envisaged a field
quantity as beable (see also Sudbery, 1987). A dissenting opinion --- one which
I do not endorse --- as to the adequacy of such a choice {\em vis-\`{a}-vis} 
the measurement problem has been recently put forward by Saunders (s.d.).}

When Bell (1984) first proposed
such a theory, he also set out to construct a {\em dynamics} --- which due 
to the discreteness of the beable had to be stochastic --- that would preserve
for all times the `quantum equilibrium' distribution for the values of $R$:
  \be
    p_i(t)=\bra{\psi(t)}P_i\ket{\psi(t)}.
    \label{twentyfour}
  \ee
Bell's problem was thus analogous to the question posed by Deotto and Ghirardi
(1998), and which we have taken up in the previous section, of constructing a
(most general) dynamics of a certain form that respects quantum equilibrium.
Bell's (1984) treatment of dynamics was further elaborated by Vink (1993)
and generalised to the case of time-dependent beables $R(t)$ by 
Bacciagaluppi and Dickson (1997) and by Sudbery (s.d.).\footnote{This 
generalisation makes it applicable to the so-called modal interpretation of 
quantum mechanics. Sudbery (s.d.) now prefers this interpretation to an 
interpretation with a time-independent beable. See Dieks and Vermaas (1998) for
a state-of-the-art collection of papers on the modal interpretation, including
discussions of dynamics, of the problems of Lorentz invariance and, most
importantly, of empirical adequacy; for another recent overview see 
Bacciagaluppi (s.d.).} I here follow Bacciagaluppi and Dickson (1997).

If one considers a closed system, the evolution may be supposed to be Markovian
(a condition which will then generally be violated by the induced evolution
of subsystems), and the corresponding process can be canonically reconstructed
from its finite-time {\em transition probabilities}. Under the appropriate
conditions these can be recovered in turn from the {\em infinitesimal} 
transition probabilities (or {\em infinitesimal parameters}) of the process.

If $p_{ji}(t,s)$ is the transition probability {\em from} state $i$ at 
time $s$ {\em to} state $j$ at time $t$ (where $t>s$), we have: 
  \be
    p_{j}(t) - p_{j}(s) = \sum_{i}p_{ji}(t,s) p_{i}(s) - p_{ij}(t,s)
    p_{j}(s),
    \label{paper42}
  \ee
and under certain assumptions, say, that the transition probabilities be 
partially differentiable with respect to $t$, we also have
the following {\em master equation}:
  \be
    \dot{p}_{j}(t) = \sum_{i} t_{ji}(t) p_{i}(t) - t_{ij}(t) p_{j}(t),
    \label{paper44}
  \ee
where the {\em infinitesimal parameters} $t_{ji}(t)$ are in fact the partial 
derivatives $\frac{\partial}{\partial t_1}p_{ji}(t_1,t_2)\vert_{t_1,t_2=t}$.

Now we can write (\ref{paper44}) in analogy to   
(\ref{six}) as a {\em continuity equation} for $p_j(t)$:
  \be
    \dot{p}_j = \sum_i j_{ji},
    \label{paper47}
  \ee
where we have defined
  \begin{equation}
    j_{ji}:= t_{ji}p_i - t_{ij}p_j.
    \label{paper45}
  \end{equation}
Thus, given that $\dot{p}_j$ is known (by the Schr\"{o}dinger equation and
(\ref{twentyfour})), we can solve the linear system of equations (\ref{paper47})
for $j_{ji}$ and then (\ref{paper45}) for $t_{ji}$, and from the $t_{ji}$
construct the stochastic process.

One has that
  \be
    \dot{p}_j(t) = 2 \Im \Big[ \bra{\psi(t)} P_j H \ket{\psi(t)} \Big],
    \label{paper70}
  \ee
and the general solution of (\ref{paper47}) is
  \begin{eqnarray}
    j_{ji}(t)  & = & 2 \Im \Big[ \bra{\psi(t)} P_j H P_i \ket{\psi(t)} \Big]+
                     \nonumber   \\[1ex]
               &   & + \bra{\psi(t)}\Big[\dot{P}_j(t)P_i(t)
                     - \dot{P}_i(t)P_j(t)\Big]\ket{\psi(t)} +
                     \nonumber   \\[1ex]
               &   & + j_{ji}^{\scriptsize \rm DG}(t),
    \label{new14}
  \end{eqnarray}
where the first term is Bell's choice of current, the second term is needed 
for time-dependent $R(t)$, and the third term is a `Deotto--Ghirardi' current
satisfying
  \be
    \sum_i j_{ji}^{\scriptsize \rm DG}(t)=0.
    \label{new15}
  \ee

Once one chooses a current, one has a further freedom in choosing the 
infinitesimal parameters $t_{ji}(t)$ in the solution of (\ref{paper45}). 
Bell's (1984) own choice was
  \be
    t_{ji} := \left\{
    \begin{array}{ll}
    {\displaystyle \frac{j_{ji}}{p_i}} \ \ & \ \ \mbox{for} \ j_{ji} > 0,
    \\[2ex]
    0 & \ \ \mbox{for} \ j_{ji} \leq 0.
    \end{array}\right.
    \label{paper99}
  \ee
Any other solution takes the form:
  \be
    t_{ji}\geq\frac{j_{ji}}{p_i}\quad\quad\quad\quad\quad\mbox{for\ }j_{ji}>0,
    \label{twentyseven}
  \ee
  \be
    t_{ji}=\frac{t_{ij}p_j-j_{ij}}{p_i}\quad\mbox{for\ }j_{ji}<0,
    \label{twentyeight}
  \ee
and
  \be
    t_{ji}=0\quad\quad\mbox{for\ }j_{ji}=0.
    \label{twentynine}
  \ee 
 
As usual, the dynamics becomes singular whenever $p_i(t)=0$ and the standard
existence and uniqueness theorem (Feller, 1940) does not apply. 
But again one can show that under the appropriate conditions the states with
zero quantum probability cannot be reached. A full proof will be given 
elsewhere, but see Bacciagaluppi (s.d., Chapter 4) for a sketch.


Both Sudbery (1987) and Vink (1993) have addressed the question of the 
continuous limit of BBB theories. Using appropriate limiting procedures 
one indeed recovers both de Broglie--Bohm theory and, as Vink shows, 
Nelson's mechanics (Sudbery remarks that there are seemingly 
sensible limiting procedures for which this is not true).

I wish to emphasise that such a dynamics has applications beyond 
the interpretational framework of beable theories.
In particular, as mentioned by Bacciagaluppi and Dickson (1997), it can be 
applied to the evolution of the {\em decohering variables} in any discrete 
model of decoherence or whenever else one considers effective
or strict {\em discrete superselection rules}.

\section{Convergence to Equilibrium and 
Experimental Constraints}\label{distribution}
In this brief note, I have reviewed the derivations of dynamics compatible 
with the assumption of the quantum equilibrium distribution, in the contexts 
of deterministic 
and stochastic configuration-space theories and of beable theories. In 
de Broglie--Bohm theory this assumption is famously known as the 
{\em distribution postulate}, and a vigorous debate has raged over the 
need and means of its justification (Berndl, Daumer, D\"{u}rr, Goldstein, 
and Zangh\`{\i}, 1995; D\"{u}rr, Goldstein, and Zangh\`{\i}, 1992; Valentini, 
1991, 1996; see also Barrett, 1995). One particular strategy has been that 
of modifying the de Broglie--Bohm theory to include, effectively
or fundamentally, a stochastic
element {\em \`{a} la} Nelson (Bohm, 1953; Bohm and Vigier, 1954; Bohm and
Hiley, 1989). In fact, in a theory with stochastic dynamics, it makes perfect 
sense to say that the epistemic distribution over the states of the system
changes in time, and in fact may approach asymptotically a distribution 
independent of any initial distribution (such distributions are known as 
{\em ergodic} distributions in the theory of Markov chains, especially 
homogeneous ones; see \eg Fisz (1963, p. 256)). Whether or not this asymptotic
behaviour is achieved depends crucially on the `mixing' properties of the
dynamics, as is very clear in Bohm and Vigier (1954, p.~211), who base their 
(at best unrigorous) derivation precisely on this assumption. Such asymptotic
behaviour has been rigorously shown to hold in Nelson's stochastic mechanics 
in the case
of the one-dimensional harmonic oscillator (Cufaro Petroni and Guerra, 1995),
and more generally for any one-dimensional system in a bound state of a
Hamiltonian of the form
  \be
    H=-\frac{\hbar^2}{2m}\frac{\partial^2}{\partial x}+V(x),
  \ee 
with time-independent $V(x)$ (Cufaro Petroni, De Martino and De Siena, s.d.). 
(There are systems for which it provably does not hold, for instance the free 
particle (Cufaro Petroni and 
Guerra, 1995).) A detailed discussion of the distribution postulate in the
BBB theories will be given in Bacciagaluppi and Barrett (s.d.). 
Here I wish to remark the following.

In Nelson's stochastic mechanics, as we have seen, the drift velocity
${\bf b}$ is made up of a current velocity, which is
equal to the de Broglie--Bohm velocity, and of an osmotic velocity
of the form
  \be
    {\bf u}=\frac{\hbar}{2m}\frac{\bfnabla\abs{\psi}^2}{\abs{\psi}^2}.
    \label{new17}
  \ee
The form of ${\bf u}$ makes it intuitively plausible that particles will be,
indeed, driven {\em away} from regions where $\abs{\psi}^2$ is small, in a way
as to achieve (asymptotic) convergence to the quantum equilibrium distribution
(cf.\ Bohm and Hiley, 1989, p. 103).

While this is obviously a desirable property of the dynamics, one might be 
suspicious of the fact that postulating an osmotic velocity {\em of the form}
(\ref{new17}) is in fact an {\em ad hoc} manoeuvre designed to favour 
precisely the convergence to $\abs{\psi}^2$. This worry now seems unjustified.
In fact, we have shown that up to an arbitrary Deotto--Ghirardi term the form 
of ${\bf u}$ follows already
from the weaker requirement that the diffusion process {\em preserve} the
quantum equilibrium distribution $\abs{\psi}^2$, and any additional velocity 
fields would presumably further
enhance any mixing properties of the dynamics.  

From this point of view, the difference between the quantum and classical 
level lies in the different ability of the dynamics to mix at different scales
(see the remarks in Bacciagaluppi, s.d., Chapters 4 and 5).
While on the sub-quantum level one requires from the dynamics that it satisfy 
assumptions of some ergodic theorem, mixing behaviour must be confined to
within the support of the {\em effective} wave function, so as to allow for 
experimental records to be permanent --- or for Schr\"{o}dinger's cat to rest
in peace. This is very welcome, since it means that there are 
situations in which one need not worry about proving the `metric 
indecomposability' of the configuration space under the (time-dependent)
quantum equilibrium measure: on the contrary, decomposability is
{\em necessary} in order to recover empirical predictions.

This brings me to my final point, namely to the idea that there are
{\em experimental constraints}
on the choice of Deotto--Ghirardi currents, whether in the deterministic or
stochastic configuration-space setting or in the beable setting. 
As in the case of the observed stability of experimental records, any such
constraints must come from the consideration of {\em individual} systems
(since any dynamics of the kind considered is by construction compatible 
with the observable statistics of ensembles), the prime example of such
systems being ions in traps.

In particular, Sudbery (s.d.) has shown that Bell's
choice of current and infinitesimal parameters correctly describes 
spontaneous decay and correctly prevents spontaneous excitation, and --- more 
strikingly --- correctly predicts the statistics of bright and dark periods of 
the `quantum telegraph', or intermittent fluorescence phenomenon
(Dehmelt, 1975; see Plenio and Knight (1998) for a recent review).
Experiments such as the quantum telegraph clearly put 
experimental constraints on the (directly exhibited) dynamics of the `hidden'
quantities --- in fact, going beyond the quantum mechanical predictions for 
ensembles, they represent novel test cases for any approach to the foundations 
of quantum mechanics.

\section*{Acknowledgements}
I am grateful to Gian Carlo Ghirardi for a copy of his talk given at the 1997
Florence--Stanford Meeting on Foundation of Physics, to Giulio Peruzzi 
for an early copy of Peruzzi and Rimini (1996). I would like to thank 
Harvey Brown, Keith Hannabuss, Lucien Hardy and Antony Valentini for 
discussion and comments, as well as audiences
at the University of Oxford and at the London School of Economics. This work 
was supported by a British Academy Postdoctoral Fellowship.

\section*{References}

\noindent
Bacciagaluppi, G. (s.d.) {\em The Modal Interpretation of Quantum
Mechanics} (Cambridge: Cambridge University Press), forthcoming.

\noindent
Bacciagaluppi, G., and Barrett, J. (s.d.), `How to Weaken
the Distribution Postulate in Pilot-Wave Theories', in preparation.

\noindent
Bacciagaluppi, G. and Dickson, M. (1997), `Dynamics for Density-Operator
Interpretations of Quantum Theory', quant-ph/9711048.



\noindent
Barrett, J. (1995), `The Distribution Postulate in Bohm's Theory', 
{\em Topoi\/} {\bf 14}, \mbox{45--54}.
 
\noindent
Bell, J. S. (1984), `Beables for Quantum Field Theory', in
Bell (1987), \mbox{pp. 173--180}.

\noindent
Bell, J. S. (1987), {\em Speakable and Unspeakable in Quantum Mechanics}
(Cambridge: Cambridge University Press).

\noindent
Berndl, K. (1996), `Global Existence and Uniqueness of Bohmian Trajectories',
in Cushing, Fine and Goldstein (1996), \mbox{pp. 77--86}.

\noindent
Berndl, K., Daumer, M., D\"{u}rr, D., Goldstein, S. and Zangh\`{\i}, N.\ 
(1995), `A Survey of Bohmian Mechanics', {\em Il Nuovo Cimento\/} 
{\bf 110 B}, \mbox{737--750}.  

\noindent
Berndl, K., D\"{u}rr, D., Goldstein, S., Peruzzi, G. and
Zangh\`{\i}, N. (1995), `On the Global Existence of Bohmian Mechanics',
{\em Communications in Mathematical Physics\/}
{\bf 173}, \mbox{647--673}.

\noindent
Bohm, D. (1952), `A Suggested Interpretation of the Quantum
Theory in Terms of ``Hidden'' Variables, I and II', {\em Physical
Review\/} {\bf 85}, \mbox{166--179}, \mbox{180--193}. 

\noindent
Bohm, D. (1953), `Proof that Probability Density Approaches $\abs{\psi}^2$
in Causal Interpretation of the Quantum Theory', {\em Physical Review\/}
{\bf 89}, \mbox{458--466}.

\noindent
Bohm, D. and Hiley, B. J. (1989), `Non-Locality and Locality in the
Stochastic Interpretation of Quantum Mechanics', {\em Physics Reports\/}
{\bf 172}, \mbox{93--122}.

\noindent
Bohm, D. and Hiley, B. J. (1993), {\em The Undivided Universe: An
Ontological Interpretation of Quantum Theory} (London: Routledge).

\noindent 
Bohm, D. and Vigier, J.-P. (1954), `Model of the Causal
Interpretation of Quantum Theory in Terms of a Fluid with Irregular
Fluctuations', {\em Physical Review} {\bf 96}, \mbox{208--216}.


\noindent
Broglie, L. de (1928), `La nouvelle dynamique des quanta', in 
{\em \'{E}lectrons et photons. Rapports et discussions du cinqui\`{e}me 
conseil de physique Solvay} (Paris: Gauthier-Villars), 
\mbox{pp. 105--132}.



\noindent
Bub, J. (1997), {\em Interpreting the Quantum World} (Cambridge: Cambridge
University Press).

\noindent
Carlen, E. (1984), `Conservative Diffusions', {\em Communications in 
Mathematical Physics\/} {\bf 94}, \mbox{293--315}.

\noindent
Cufaro Petroni, N., De Martino, S. and De Siena, S. (s.d.),
`Exact Solutions of Fokker--Planck Equations Associated to Quantum Wave
Functions', forthcoming in {\em Physics Letters} {\bf A}.

\noindent
Cufaro Petroni, N. and Guerra, F. (1995), `Quantum Mechanical States as 
Attractors for Nelson Processes', {\em Foundations of Physics\/} {\bf 25}, 
\mbox{297--315}.

\noindent
Cushing, J. T., Fine, A. and Goldstein, S. (eds) (1996), {\em Bohmian
Mechanics and Quantum Theory: An Appraisal. Boston Studies in the
Philosophy of Science}, Vol.~184 (Dordrecht: Kluwer).

\noindent
Davidson, M. (1979), `A Generalization of the F\'{e}nyes--Nelson Stochastic
Model of Quantum Mechanics', {\em Letters in Mathematical Physics\/} {\bf 3},
\mbox{271--277}.

\noindent
Dehmelt, H. (1975), `Proposed $10^{14}\Delta\nu>\nu$ Laser Fluorescence
Spectroscopy on Tl$^+$ Mono-Ion Oscillator, II', {\em Bulletin of the American
Physical Society} {\bf 20}, 60.

\noindent
Deotto, E. and Ghirardi, G. C. (1998), `Bohmian Mechanics Revisited',
{\em Foundations of Physics} {\bf 28}, \mbox{1--30}.


\noindent
Dieks, D. and Vermaas, P. E. (eds) (1998), {\em The Modal Interpretation of
Quantum Mechanics} (Dordrecht: Kluwer). 

\noindent
D\"{u}rr, D., Goldstein, S. and Zangh\`{\i}, N. (1992), `Quantum
Equilibrium and the Origin of Absolute Uncertainty', {\em
Journal of Statistical Physics\/} {\bf 67}, \mbox{843--907}. 



\noindent
Feller, W.\ (1940), `On the Integro-Differential Equations of Purely
Discontinuous Markoff Processes', {\em Transactions of the American
Mathematical Society} {\bf 48}, \mbox{488--515}, and `Errata', {\em ibid.} 
{\bf 58} (1945), 474. 

\noindent
Fisz, M. (1963), {\em Probability Theory and Mathematical Statistics\/}
(New York: Wiley, 3rd edn).

\noindent
Gurtler, R. and Hestenes, D. (1975), `Consistency of the Dirac, Pauli and 
Schr\"{o}dinger Theories', {\em Journal
of Mathematical Physics} {\bf 16}, \mbox{573--584}.




\noindent
Holland, P. (1993), {\em The Quantum Theory of Motion} (Cambridge:
Cambridge University Press).



\noindent
Nelson, E. (1966), `Derivation of the Schr\"{o}dinger Equation from Newtonian
Mechanics', {\em Physical Review\/} {\bf 150}, \mbox{1079--1085}.

\noindent
Nelson, E. (1985), {\em Quantum Fluctuations}
(Princeton: Princeton University Press).

\noindent
Peruzzi, G. and Rimini, A. (1996), `Quantum Measurement in a Family of
Hidden-Variable Theories', {\em Foundations of Physics Letters\/} {\bf 9},
\mbox{505--519}. 

\noindent
Plenio, M. and Knight, P. L. (1998), `The Quantum-Jump Approach to Dissipative
Dynamics in Quantum Optics', {\em Reviews of Modern Physics} {\bf 70}, 
\mbox{101--144}.

\noindent
Saunders, S. (s.d.), `Relativistic Pilot-Wave Theory', forthcoming in 
J.~N.~Butterfield and C.~Pagonis (eds), {\em From Physics to Philosophy}
(Cambridge: Cambridge University Press).

\noindent
Squires, E. (1996), `Local Bohmian Mechanics', in Cushing, Fine and Goldstein 
(1996), \mbox{131--140}.

\noindent 
Sudbery, A. (1986), {\em Quantum Mechanics and the Particles of Nature}
(Cambridge: Cambridge University Press).

\noindent 
Sudbery A. (1987), `Objective Interpretations of Quantum Mechanics and the 
Possibility of a Deterministic Limit', {\em Journal of Physics A: Math.\ Gen.\ }
{\bf 20}, \mbox{1743--1750}.

\noindent
Sudbery, A. (s.d.), in preparation.

\noindent
Valentini, A. (1991), `Signal-Locality, Uncertainty, and the
Subquantum H-Theorem, I and II', {\em Physics Letters\/} {\bf A 156}, 
\mbox{5--11}, {\bf A 158}, \mbox{1--8}.

\noindent
Valentini, A. (1996), `Pilot-Wave Theory of Fields, Gravitation and
Cosmology', in Cushing, Fine and Goldstein (1996),
\mbox{pp. 45--66}.



\noindent
Vink, J. (1993), `Quantum Mechanics In Terms of Discrete
Beables', {\em Physical Review\/} {\bf A 48}, \mbox{1808--1818}.

\noindent
Wallstrom, T. (1994), `Inequivalence Between the Schr\"{o}dinger Equation and 
the Madelung Hydrodynamic Equations', {\em Physical Review\/}
{\bf A 49}, \mbox{1613--1617}.


\end{document}